\documentclass[letter]{jpsj2}
\usepackage{latexsym}
\usepackage{color}

\newcommand{\nn}{\nonumber \\}
\newcommand{\udarrow}[2]{\smash{\mathop{%
  \hbox to 0.4cm{$\rightleftharpoons$}}\limits^{#1}\limits_{#2}}}

\title{%
Chirality Selection in Crystallization
}

\author{%
Yukio \textsc{Saito}\thanks{yukio@rk.phys.keio.ac.jp}
 and Hiroyuki \textsc{Hyuga}\thanks{hyuga@rk.phys.keio.ac.jp}
}

\inst{%
Department of Physics, Keio University, Yokohama 223-8522 
}

\recdate{\today}

\abst{%
A cluster growth model is proposed to study chiral symmetry breaking in
stirred crystallization from an achiral element.
Achiral monomers are assumed to coagulate to form
 chiral clusters from dimers to hexamers.
Due to the stirring,  the hexamers break into dimers.
The coagulation of two dimers into a tetramer also occurs.
The fixed point analysis of coupled rate equations of cluster densities
and their numerical integrations show 
that the chiral symmetry becomes broken if dimers are critical and able to 
dissociate back to achiral monomers.
The chirality selection is understood 
by showing that, in a quasi-steady approximation, 
the rate equations reduce to those 
of dimers with nonlinear autocatalysis and decomposition.
}

\kword{%
homochirality, crystal growth,  enantiomer,
rate equation, dynamical chiral symmetry breaking
}

\begin{document}
\sloppy
\maketitle

It has been known for a long time that 
organic molecules in life are
homochiral with a completely broken chiral symmetry, 
and their origin has intrigued many scientists.
\cite{feringa+99}
Frank showed theoretically that an autocatalytic reaction
with cross inhibition between two enatiomers leads to the selection of
one type of enantiomer in an open system.
\cite{frank53}
We have recently proven that a system with nonlinear autocatalysis
and back reaction as is described by the following rate equations
\begin{align}
 \dot r = (k_0+ k_1 r + k_2 r^2) a - \lambda r,
\nonumber \\
 \dot s = (k_0+ k_1 s + k_2 s^2) a - \lambda s,
\label{eq1}
\end{align}
can lead to complete homochirality in a closed
system.\cite{saito+04a,saito+04b,saito+04c}
Here, $a$ denotes the concentration of an achiral substrate, $r$ and $s$ are 
those of chiral enantiomers, and the conservation relation
$a+r+s=$const. is assumed to hold.
The model is motivated by the chemical reaction system discovered by 
Soai {\it et al.},\cite{soai+95,sato+03}
 which 
is the first example of chemical reaction to show the amplification of enantiomeric excess in a closed system.

Chiral symmetry breaking, however, is found not only in biological and chemical
systems, but also
in other systems such as in crystallization.
\cite{kondepudi+01}
Even achiral inorganic molecules can gather together to form chiral crystals. 
Examples are 
a quartz of SiO$_2$ crystal, or a crystal of sodium chlorate (Na\,ClO$_3$).
When the sodium chlorate crystal grows  from a solution in a cell, 
a racemic mixture of $D$ and $L$ crystals emerges.
On the other hand, when the crystal is grown 
from a supersaturated solution
under continuous stirring,
the growth cell is occupied by one type of enantiomer.
\cite{kondepudi+90}
The chirality selection is found to be brought about
by fragmenting an initially nucleated crystal: 
the scattered clusters of small crystal act as
centers of secondary nucleation
of the same chirality as that of the initial crystal.
This senario is called a " single mother" crystallization.
Recently, however, it has been found that even if the initial solution contains
crystalline seeds of 
both enantiomers, the continuous stirring leads to 
the state of complete chiral purity with one type of enantiomer.
The selected enantiomer is the one which 
was in a slight excess initially. 
\cite{viedma04}
There is a theoretical study associated with this phenomenon,
\cite{uwaha04}
but we propose another approach in order to examine the roles of
various elementary processes.
We also study the relationship between the present model and eq. (\ref{eq1})
reported in our previous studies.
\cite{saito+04a,saito+04c}

Our model of crystallization is described in terms of clusters of
various sizes.
An individual molecule A is achiral, but 
a few of them coagulate into a chiral embryo. 
For simplicity, we assume here that dimers already
show the chirality.
We call the right-handed dimer R$_2$, and the
left-handed one S$_2$.
The first growth process is then described by
\begin{align}
A+A ~ \udarrow{k_1}{\lambda} R_2, & \quad
A+A ~ \udarrow{k_1}{\lambda} S_2 .
\label{eq2}
\end{align}
These chiral embryos grow further by incoorporating elements A 
one by one
as,
\begin{align}
R_{i} + A \stackrel{k_i}{\rightarrow} R_{i+1},
 \qquad
S_{i} + A \stackrel{k_i}{\rightarrow} S_{i+1},
\label{eq3}
\end{align}
with $i \ge 2$.
Here, we assume for simplicity that the critical size of the crystal is $i=2$.
Only the dimers can dissociate with a rate $\lambda$, which is assumed
 larger than the growth rate $k_1$ from monomers into a dimer.
Clusters 
larger than
the dimer are stable and only grow.
If these are the whole processes, 
the crystalline enantiomeric excess does not increase.
In the experiment\cite{viedma04} the solution is continuously stirred by
a stirrer and balls,\cite{viedma04} 
which crash part of the grown crystals and split off small clusters.

First we consider the case that
the smallest nontrivial cluster, tetramer, splits as
\begin{align}
R_4 \stackrel{\nu}{\rightarrow} R_2 +R_2,
\qquad
S_4 \stackrel{\nu}{\rightarrow} S_2 +S_2
\label{eq4}
\end{align}
with a rate $\nu$.
Coupled rate equations for concentrations 
$r_{i},~s_{i}$ of dimers ($i=2)$, trimers ($i=3$) and tetramers ($i=4$) 
are written as
\begin{align}
&\dot{r}_2=k_1a^2-\lambda r_2-k_2r_2a+2\nu r_4\nn
&\dot{r}_3=k_2r_2a-k_3r_3a\nn
&\dot{r}_4=k_3r_3a-\nu r_4\
\label{eq5}
\end{align}
and the corresponding ones for S's.
Since the total amount of the molecules is conserved in a closed reaction
vessel, the concentration of elementary molecule A is given as
$a=c_0-2(r_2+s_2)-3(r_3+s_3)-4(r_4+s_4)$ with a constant $c_0$.
Numerical integration indicates readily that the chiral symmetry 
will not be broken in the present case.
This result can be easily undestood by applying the 
quasi-steady state approximation.
Because of large reaction rates $k_i ~( i \ge 2)$ and $\nu$,
clusters 
larger than the dimers
are assumed  to adjust their concentrations 
rapidly to their steady-state values as
$r_3= (k_2/k_3) r_2,~r_4=(k_2/\nu) r_2a$, 
as long as $a \ne 0$.
Then the evolution of the dimer concentrations, $r_2$ and $s_2$,
 is simply described by the reduced rate equations
\begin{align}
\dot{r}_2=k_1a^2-\lambda r_2+k_2ar_2 ,
\nonumber \\
\dot{s}_2=k_1a^2-\lambda s_2+k_2as_2 .
\label{eq6}
\end{align}
The equation contains only a linear autocatalytic process and a back reaction,
and looks similar to eq. (\ref{eq1}) with $k_2=0$.
\cite{saito+04a} 
A simple calculation shows that there is only a racemic 
fixed point as 
$r_2=s_2=k_1 a^2/(\lambda - k_2 a)$. Therefore, the
system approaches a racemic state asymptotically, in agreement with
numerical integration results.
 The  chiral symmetry is conserved.
The key point is that the secondary nucleation induces the autocatalytic effect
in the dimer evolution, but only up to the level of linear autocatalysis.
Note that the fragmentation effect $\nu$ does not 
appear explicitly in the reduced rate equation (\ref{eq6}).

Since the chirality is not selected in the cluster 
growth up to the tetramer, we go one step further, up to the hexamer;
 R$_i$, S$_i$ with $2 \le i \le 6$.
Instead of eq. (\ref{eq4}), we assume the breaking of a hexamer into
three dimers as
\begin{align}
R_6 \stackrel{\nu}{\rightarrow} 3 R_2 ,
\qquad
S_6 \stackrel{\nu}{\rightarrow} 3 S_2 ,
\label{eq7}
\end{align}
and furthermore the incorporation of the coagulation of dimers into
a tetramer
\begin{align}
R_2 +R_2 \stackrel{\mu}{\rightarrow} R_4  ,
\qquad
S_2 +S_2 \stackrel{\mu}{\rightarrow} S_4
\label{eq8}
\end{align}
Then the rate equations are written as 
\begin{align}
&\dot{r}_2=k_1a^2-\lambda r_2-k_2r_2a-2\mu r_2^2+3\nu r_6\nn
&\dot{r}_3=k_2r_2a-k_3r_3a\nn
&\dot{r}_4=k_3r_3a-k_4r_4a+\mu r_2^2\nn
&\dot{r}_5=k_4r_4a-k_5r_5a\nn
&\dot{r}_6=k_5r_5a-\nu r_6
\label{eq9}
\end{align}
and the corresponding ones for S's, under the conservation condition
$a=c_0-\sum_{i=2}^6 i(r_i+s_i)$.

Since these coupled  equations involve many degrees of freedom, 
detailed analysis can be performed only by numerical integration.
However, the essential feature may be understood 
in a quasi-steady approximation as follows.
By assuming the rapid variation of concentrations of
clusters larger than the critical size $i=2$,
their concentrations are approximated by 
those values determined by the given concentrations of the critical cluster as;
$r_3=(k_2/k_3)r_2,~r_4=(k_2/k_4)r_2+(\mu/k_4a)r_2^2,~
r_5=(k_2/k_5)r_2+(\mu/k_5a)r_2^2,~r_6=(k_2/\nu)r_2a+(\mu/\nu)r_2^2$,
and the corresponding relations for $s$'s, as long as $a \ne 0$.
In particular, we obtain the relation $\nu r_6 = k_2r_2a+\mu r_2^2$.
By inserting the relation into the rate equations of the dimers, one obtains
the reduced rate equations for the dimers as
\begin{align}
\dot r_2 = k_1a^2 - \lambda r_2 +2k_2r_2a + \mu r_2^2,
\nonumber \\
\dot s_2 = k_1a^2 - \lambda s_2 +2k_2s_2a + \mu s_2^2.
\label{eq10}
\end{align}
These equations contain essential terms such as the decomposition term 
$-\lambda r_2$ (or $- \lambda s_2)$ and a nonlinear autocatalytic term 
$\mu r_2^2$ ( or $\mu s_2^2)$.
In this sense we expect that eq. (\ref{eq10}) shows similar characteristics
to that of eq. (\ref{eq1}).
For example, if the dissociation is absent, $\lambda=0$, then 
$\dot r_2$ and $\dot s_2$ consist only of positive terms and
the stationary solution $\dot r_2= \dot s_2=0$ is possible only for
$a=r_2=s_2=0$. 
From the stationary condition for hexamers we obtain $r_6=s_6=0$, 
but other concentrations
$r_i$ and s$_i$ with $3 \le i \le 5$ are arbitrary as long as 
the conservation relation
is satisfied. No chirality selection is possible, since there are no fixed 
points, but fixed lines or planes. 

We now study the case with dissociation ($\lambda >0$).
One notices that the fixed points for the full rate equation (\ref{eq9})
 are 
identical with those obtained by the reduced rate equation (\ref{eq10})
by way of the construction of the latter.
There are racemic fixed points ($r_2= s_2)$ which are not of our interest.
As for the asymmetric fixed points, it can be easily shown that they 
should satisfy the relations
\begin{align}
&r_2+s_2=\frac{\lambda-2k_2a}{\mu}>0 ,
\qquad
r_2s_2 = \frac{k_1a^2}{\mu },
\nonumber \\
&c_0=a+(A+Ba)k_2(r_2+s_2)+(C/a+B)\mu(r_2^2+s_2^2),
\label{eq11}
\end{align}
with 
$A=\sum_{i=2}^5 ik_i^{-1}, B=6/\nu , 
C=\sum_{i=4}^5 ik_i^{-1} .
$
The first two relations of eq. (\ref{eq11}) yield two asymmetric fixed points 
S$_1$:$(r_{2,+},~s_{2,-})$ and S$_2$:$(r_{2,-},~s_{2,+})$ with
\begin{align}
&r_{2,\pm}=\frac{\lambda-2k_2a}{2\mu}\pm 
\sqrt{\left(\frac{\lambda-2k_2a}{2\mu}\right)^2-\frac{k_1a^2}{\mu}}
=s_{2,\mp}  ,
\label{eq12}
\end{align}
which is expressed by using 
a fixed point value of $a$ yet to be determined.
For the existence of these asymmetric fixed points, 
$a$ should be in the range
$
{\lambda}/{2(k_2+\sqrt{\mu k_1})}\ge a>0. 
$ 
Finally the asymmetric fixed points are determined by solving an algebraic
equation of $a$, 
obtained by inserting the first two relations in eq. (\ref{eq11}) 
into the last one. 

Since the conditions for the existence of the asymmetric fixed points are 
very involved, 
we restrict the discussion here to the case of $c_0=k_2=k_3=k_4=k_5=1$. 
Then it can be shown
that the asymmetric fixed points exist for $\lambda$'s in the range
\begin{align} 
0<\lambda<\lambda_u ,
\label{eq13}
\end{align}
where the upper limit $\lambda_u$ is given by the larger root 
of the quadratic equation
\begin{align}
6\beta \lambda^2+\nu(\mu+28\beta+18\beta^2)\lambda-2\mu\nu(1+\beta)=0
\label{eq14}
\end{align}
with $\,\beta=\sqrt{\mu k_1}\,$. 
In particular, $\lambda_u \simeq 2$ for $k_1\simeq 0\,$ 
and $\lambda_u \simeq \mu\nu/9\beta\rightarrow 0 $ 
for $k_1\rightarrow \infty\,$. 

We then perform numerical integration of the rate equation (\ref{eq9}),
in order to observe the chiral symmetry breaking in the crystallization.
Since the formation of a critical nucleus (= a dimer) is expected to be slow,
 $k_1$ is set as small as $k_1=0.01$
compared with other parameters as 
$c_0=1,~k_2=k_3=k_4=k_5=1,~\lambda=0.3,~\mu=\nu=1$.
The decomposition rate $\lambda=0.3$ is set smaller than the upper limit
$\lambda_u=0.513$ calculated from eq. (\ref{eq14}), and 
one of the asymmetric fixed points is expected to be realized.
This expected behavior is in fact observed in the flow diagram shown in Fig. 1
in the phase space of the total number densities of
R and S crystals, $r_{_{\Sigma}}=\sum_{i=2}^6 ir_i$ and 
$s_{_{\Sigma}}=\sum_{i=2}^6 is_i$, respectively.
By starting close to the symmetric mixture 
$r_{_{\Sigma}} \approx s_{_{\Sigma}}$,
both enantiomeric crystals increase in number, approaching the
racemic fixed point, but this racemic fixed point is unstable, and eventually
one enantiomer overwhelms the other.
The flow is attracted to one of the chiral fixed points.
If one decreases $k_1$, the chiral asymmetry represented by
the crystalline enantiomeric excess cee
 $=|(r_{_{\Sigma}}- s_{_{\Sigma}})
/ (r_{_{\Sigma}}+ s_{_{\Sigma}})|$
increases up to almost unity.

\begin{figure}[h]
\begin{center} 
\includegraphics[width=0.75\linewidth]{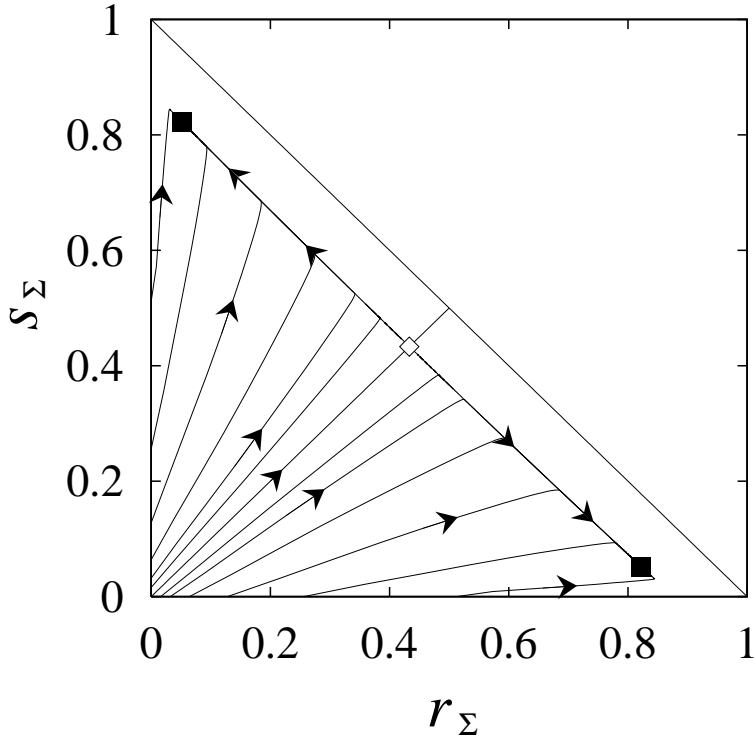}
\end{center} 
\caption{Flow diagram in total $r$'s and $s$'s phase space.
Symbols represents symmetric and asymmetric fixed points. 
The parameters are 
$c_0=1,~k_1=0.01,~k_2=k_3=k_4=k_5=1,~\lambda=0.3,~\mu=\nu=1$.
}
\label{fig1}
\end{figure}

To summarize, 
we proposed a clustering model for the growth of a chiral crystal 
composed of achiral elements, and studied the
conditions under which chiral symmetry breaking takes place.
The dimers are assumed to be the smallest size  
possessing chiralities, and at the same time of
the critical size.
When the cluster size increases indefinitely by successive
incorporation of achiral elements, a racemic mixture results.

In experiments, chirality selection is achieved by continuous stirring such
that large crystals are fragmented into small clusters
and the scattered small clusters induce secondary nucleation.
In order to simulate the effect of the fragmentation caused by stirring, 
we introduce a decomposition process of large clusters 
into dimers in our crystallization model. 
If we restrict the cluster growth only up to the tetramers, 
the chiral symmetry is never broken.
The result is interpreted in the quasi-steady approximation 
where clusters larger than the critical size are assumed to
evolve sufficiently rapidly to attain the steady-state concentration. 
The approximation
leads to the reduced rate equation of dimers with a linear autocatalysis,
which is shown to be insufficient
for breaking the chiral symmetry, as reported in our previous study.
\cite{saito+04a}

To equip the the model with the ability to break chiral symmetry dynamically,
one needs a nonlinear autocatalytic process in the reduced rate equation
 of the dimers.
This is acomplished by incorporating dimers
into the other clusters to facilitate cluster growth.
The process leads effectively to the nonlinear enhancement 
of the dimer production rate,
or a nonlinear autocatalytic effect. 
We know from our previous study that the chiral symmetry can break
in appropriate situations.
\cite{saito+04a,saito+04b,saito+04c}
 Numerical integration of the full rate equations
confirms this expectation.
In the present model we included the coagulation of two dimers into a tetramer.
This process consumes the minimum chiral units or dimers in a nonlinear way,
but the fragmentation of a larger cluster, hexamer in our present case,
replenishes dimers again into the crystallization process.
If we incorporate dimer coagulation with clusters of larger sizes
as R$_2$+R$_j \rightarrow $R$_{j+2}$ and the breaking
processes R$_{j+2} \rightarrow $R$_2$+R$_j$, the
parameter region 
for the chiral symmetry breaking is affected
but the qualitative feature remains valid.
In the present model, the dissociation which corresponds 
to the back reaction 
is again found essential, since
without it the system relaxes to the fixed line in 
$r_{_{\Sigma}} - s_{_{\Sigma}}$ phase space, and no 
precise chirality selection takes place. 

Thus far we have assumed that the smallest clusters, dimers, which show 
chirality
are at the same time critical clusters which can dissociate, 
for mathematical simplicity.
Actually there should be a margin between the two sizes, and
the fragmentation of large clusters into those with sizes smaller than the 
critical ones might be the minimal requirement to sustain symmetry breaking.
This aspect will be studied in the future. 

\acknowledgement
Authors acknowledge support from the Gakuji-Shinkou-Shikin 
by Keio University.


\end{document}